**A generalized approach to photon avalanche upconversion in luminescent nanocrystals**

Artiom Skripka[1,2*], Minji Lee[1,3], Xiao Qi[1], Jia-Ahn Pan[1], Haoran Yang[1,#], Changhwan Lee[4], P. James Schuck[4], Bruce E. Cohen[1,5], Daniel Jaque[2*] and Emory M. Chan[1*]

[1]The Molecular Foundry, Lawrence Berkeley National Laboratory, Berkeley, California 94720, United States

[2]Nanomaterials for Bioimaging Group, Departamento de Física de Materiales, Facultad de Ciencias, Universidad Autónoma de Madrid, Madrid, 28049, Spain

[3]Department of Chemical and Biomolecular Engineering, University of California Berkeley, Berkeley, California 94720, United States

[4]Department of Mechanical Engineering, Columbia University, New York, New York 10027, United States

[5]Division of Molecular Biophysics & Integrated Bioimaging, Lawrence Berkeley National Laboratory, Berkeley, California 94720, United States

[#]Present address: STMicroelectronics, Santa Clara, California, United States

*Corresponding authors: a_skripka@lbl.gov, daniel.jaque@uam.es, emchan@lbl.gov

**Abstract:** Photon avalanching nanoparticles (ANPs) exhibit extremely nonlinear upconverted emission valuable for sub-diffraction imaging, nanoscale sensing, and optical computing. Avalanching has been demonstrated with $Tm^{3+}$, $Nd^{3+}$ or $Pr^{3+}$-doped nanocrystals, but their emission is limited to 600 and 800 nm, restricting applications. Here, we utilize $Gd^{3+}$-assisted energy migration to tune the emission wavelengths of $Tm^{3+}$-sensitized ANPs and generate highly nonlinear emission of $Eu^{3+}$, $Tb^{3+}$, $Ho^{3+}$, and $Er^{3+}$ ions. The upconversion intensities of these spectrally discrete ANPs scale with the nonlinearity factor $s$ = 10-17 under 1064 nm excitation at power densities as low as 6 kW·cm$^{-2}$. This strategy for imprinting avalanche behavior on remote emitters can be extended to fluorophores adjacent to ANPs, as we demonstrate with CdS/CdSe/CdS core/shell/shell quantum dots. ANPs with rationally designed energy transfer networks provide the means to transform conventional linear emitters into a highly nonlinear ones, expanding the use of photon avalanching in biological, chemical, and photonic applications.

Photon avalanching (PA) nanoparticles exhibit giant optical nonlinearities in their upconverted emission, such that a 10-20% increase in near infrared (NIR) excitation power can result in a 100- to 1000-fold increase in the emission intensity.[1] PA behavior has been recently reported at room temperature in lanthanide ($Ln^{3+}$)-doped nanocrystals of $NaYF_4:Tm^{3+}$,[2] $LiYF_4:Tm^{3+}$, $KPb_2Cl_5:Nd^{3+}$, and $NaYF_4:Yb^{3+}, Pr^{3+}$.[3–5] Their highly nonlinear responses, equivalent in some cases to those of >30-photon processes, has been utilized for nanoscale thermometers, molecular rulers, and most saliently, in sub-diffraction microscopy[6–10] to achieve optical resolutions finer than 100 nm using conventional laser scanning microscope.[2] However, these early avalanching nanoparticles (ANPs) mostly emit in the NIR (i.e., 800 nm), which limits the ability to probe multiple species independently at different visible wavelengths. To realize multiplexed applications of PA in imaging, sensing, and computing, ANPs with a broader wavelength range and more spectrally distinct luminescence profiles must be developed.

To diversify the library of ANPs using the few proven PA compositions, we utilized energy migration to transfer the nonlinear behavior of the $Ln^{3+}$ ions responsible for avalanching to physically segregated emitters within the same nanocrystals and also to neighboring nanoparticles. Similar energy migration upconversion (EMU) approaches have been used in conventional upconverting nanoparticles (UCNPs) to tune upconversion with various $Ln^{3+}$ emitters without changing underlying light sensitization mechanisms.[11] Such physical decoupling is even more essential for ANPs, since the energy looping mechanism underpinning PA is particularly susceptible to quenching via cross-relaxation when additional ions are co-doped with avalanching ions. In ANPs, energy migration through $Yb^{3+}$ ions was recently used to transfer the avalanching behavior of the $Yb^{3+}/Pr^{3+}$ PA system to ions that emit at visible wavelengths.[5] This approach, however requires high excitation powers (>70 kW·cm$^{-2}$) and deconvolution of overlapping emission signals.

Here, we demonstrate how PA originating from $Tm^{3+}$ ions can be harnessed together with EMU through $Gd^{3+}$ ions to create a library of spectrally discrete ANPs with low avalanching thresholds and highly nonlinear emissions (**Figure 1**). We posited that reducing the number of avalanching $Tm^{3+}$ ions (< 500) in small $NaGdF_4$ core nanoparticles (< 7 nm) would accelerate saturation of the $^3H_4$ state and facilitate population of higher-energy levels of $Tm^{3+}$ ($^1D_2$ and $^1I_6$).[8,12] Subsequently, energy from these excited $Tm^{3+}$ ions can be transferred to $Gd^{3+}$ ions in the host matrix and relayed to any activator lanthanide ion ($A^{3+}$) in a nanocrystal. Because the energy of $Gd^{3+}$ ions ($^6P_J$) is transferred from $Tm^{3+}$ ions excited by multiple stages of PA, the high nonlinearity of PA can be preserved in the emission of $A^{3+}$ ions (**Figure 1b**). Furthermore, EMU in ANPs can be extended to semiconductor quantum dots (QD), prompting highly nonlinear emission from an

ANP+QD avalanching complexes. Such avalanching complexes could enable development of long-range energy transfer sensors with high localization precision.[9] Overall, the combination of $Tm^{3+}$-PA and $Gd^{3+}$-EMU provides a way to easily customize upconversion emission of ANPs, and the demonstrated spectrally discrete ANPs and avalanching complex enrich the palette of available nonlinear emitters for a range of photonic applications.

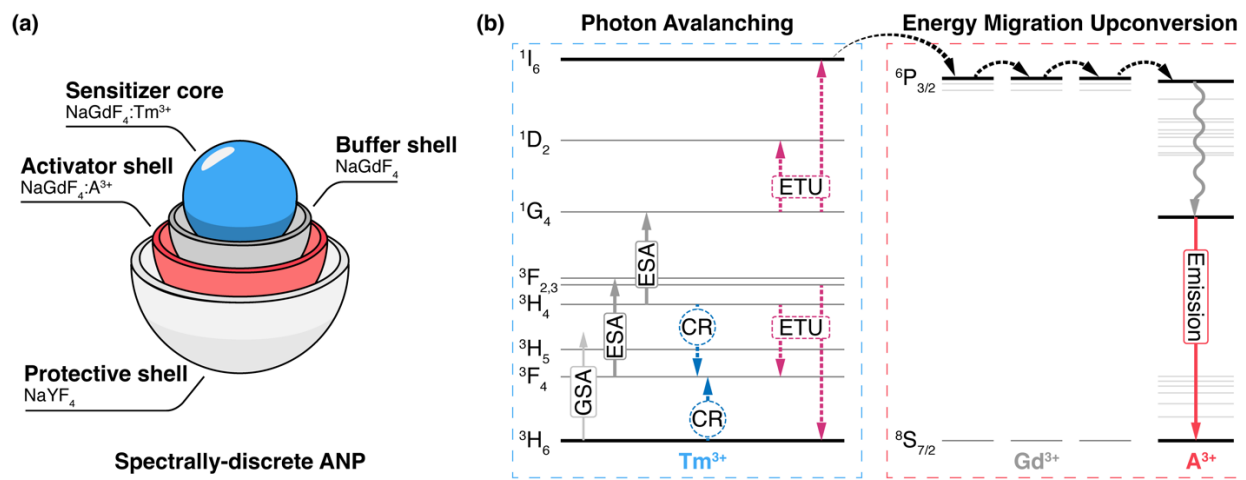

**Figure 1**. (a) Schematic representation of spectrally discrete ANPs consisting of the sensitizer core ($NaGdF_4:Tm^{3+}$), buffer shell ($NaGdF_4$), activator shell ($NaGdF_4:A^{3+}$), and the protective shell ($NaYF_4$). (b) Simplified energy level scheme and operation of spectrally discrete ANPs depicting 1064-nm-excited photon avalanche (GSA/ESA – ground/excited state absorption, CR – cross-relaxation, ETU – energy transfer upconversion), energy migration via $Gd^{3+}$ ions, and instigation of visible avalanche upconversion in $A^{3+}$ ions. Solid arrows – photon emission/absorption, dashed arrows – non-radiative energy transfer, wavy arrow – non-radiative multiphonon relaxation.

To implement this EMU photon avalanching architecture, we synthesized heterostructured ANPs consisting of a $NaGdF_4$: 20 mol% $Tm^{3+}$ sensitizer core, an undoped $NaGdF_4$ buffer shell, a $NaGdF_4$: 15 mol% $Eu^{3+}$ activator shell, and an undoped $NaYF_4$ outer shell (**Figure 2a, b**). Each shell of ANPs was grown sequentially by thermal decomposition of precursors using a nanoparticle synthesis robot (WANDA, see **Supporting Information S1** for details).[13,14] An intermediate $NaGdF_4$ buffer shell was introduced to minimize direct energy transfer between sensitizer ($Tm^{3+}$) and activator ($Eu^{3+}$) ions, which could lead to quenching.[15] Meanwhile, the protective $NaYF_4$ outer shell (**Figure S1** shows elemental mapping of Gd and Y) was used to reduce surface quenching and prevent excitation energy spill-out from the $Gd^{3+}$ network.[16] As synthesized, the $Eu^{3+}$-activated ANPs were observed to be pure $Na(Y/Gd)F_4$ β-phase (**Figure**

**S2**) and less than 25 nm in diameter, determined by powder X-ray diffraction (XRD) and transmission electron microscopy (TEM), respectively. High crystallinity, small-size, and narrow size distribution (< 5%) ideally positions these ANPs for biological or nanoscale patterning applications.[17–19]

Under 1064 nm irradiation the $Eu^{3+}$-activated ANPs emitted 800 nm light characteristic of $^3H_4 \rightarrow {}^3H_6$ radiative relaxation in avalanching $Tm^{3+}$ ions (**Figure S3**).[2,13] Interestingly, we observed a series of visible emission peaks at 510, 535, 555, 580, 590, and 615 nm, consistent with $Eu^{3+}$ radiative transitions from $^5D_J$ excited states (**Figure 2c**). The presence of $Eu^{3+}$ emission confirms that energy migration from the $Tm^{3+}$-doped core to the $Eu^{3+}$-doped shell occurs through the $Gd^{3+}$ sublattice, especially since no emission was detected with $Tm^{3+}$-free core/shell $NaGdF_4$: 15 mol% $Eu^{3+}/NaYF_4$ control nanoparticles excited at 1064 nm (**Supporting Information S3**). We note that the emission profile of $Eu^{3+}$ in the 500-640 nm range has no overlap with $Tm^{3+}$ upconversion and can be readily detected without the need for spectral deconvolution.

To determine whether PA occurs in these $Eu^{3+}$-activated ANPs, we investigated the power dependence of the visible $Eu^{3+}$ emission. With pump power ($P$) increasing above the 25 kW·cm$^{-2}$ threshold, the intensity ($I$) of the $Eu^{3+}$ ($^5D_0 \rightarrow {}^7F_2$) line at 615 nm increased with the nonlinearity factor $s$ = 14.6, where $s$ is the slope of the log-log plot in **Figure 2d**, and $I \propto P^s$. This steep power scaling of $Eu^{3+}$ emission (i.e., $s$ > 10) strongly suggests that the nonlinear behavior of the avalanching $Tm^{3+}$ core is preserved in high energy states (e.g. $^1I_6$) and can be transferred to $Eu^{3+}$ ions by $Gd^{3+}$-facilitated energy migration. As an additional indication of PA in $Eu^{3+}$-activated ANPs, we observed prolonged rise times in time resolved luminescence of $Eu^{3+}$ emission and their subsequent shortening with the increasing pump power (**Figure S4**), both of which are signatures of PA.[1]

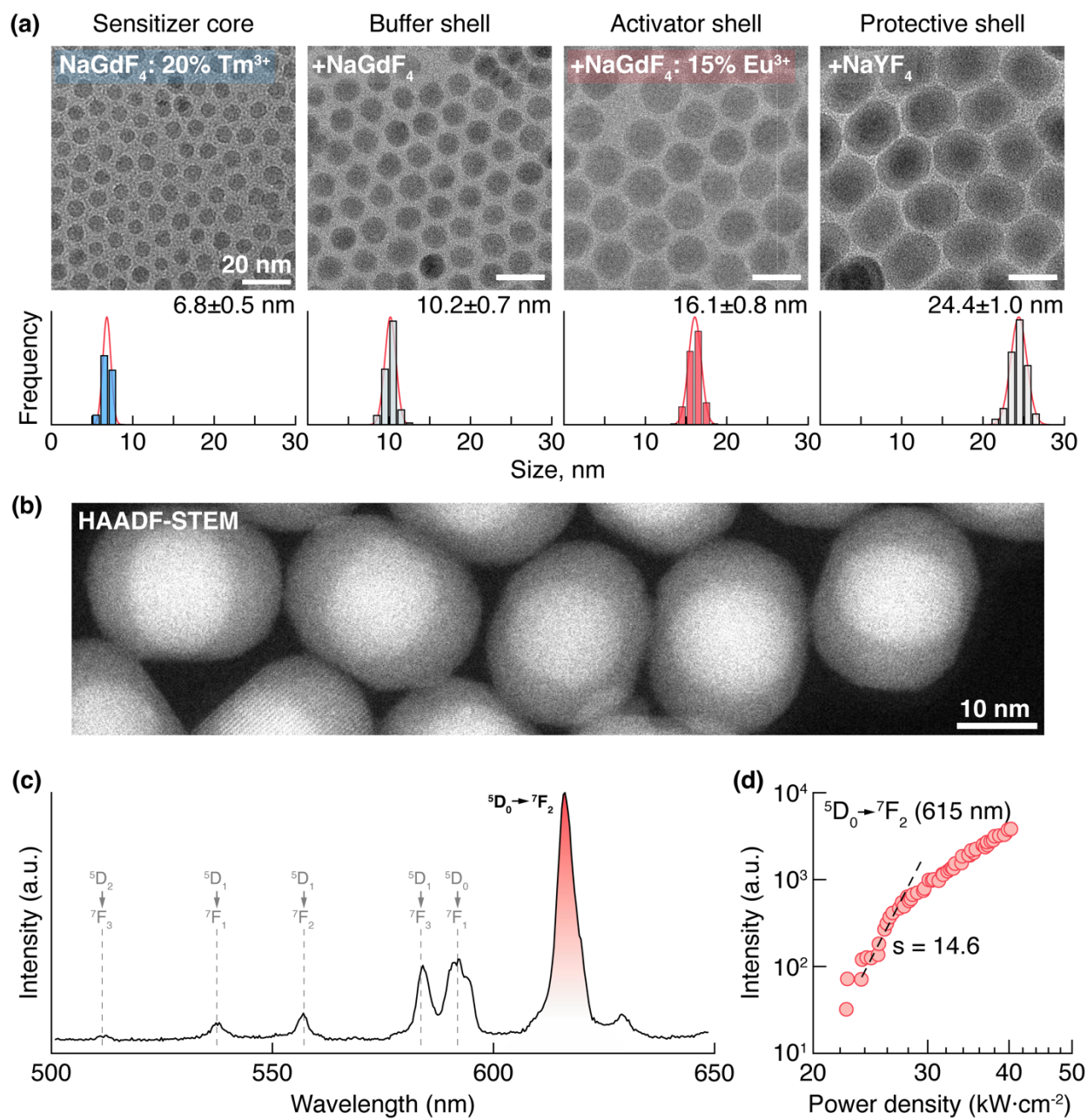

**Figure 2**. (a) Representative TEM micrographs of NaGdF$_4$: 20% Tm$^{3+}$/NaGdF$_4$/NaGdF$_4$: 15% Eu$^{3+}$/NaYF$_4$ ANPs following each synthesis step. Scale bar = 20 nm. Corresponding nanoparticle size distribution histograms together with average size and one standard size deviation are shown below each micrograph. (b) High-angle angular dark-field scanning transmission electron microscopy (HAADF-STEM) image of Eu$^{3+}$-activated ANPs. (c) Eu$^{3+}$ upconversion emission spectra in ANPs under 1064 nm excitation (49 kW·cm$^{-2}$). The identified Eu$^{3+}$ transitions are labeled next to each band. (d) Pump power dependence of Eu$^{3+}$ [$^5D_0 \rightarrow {}^7F_2$; red-highlight in (c)] emission

under 1064 nm excitation. Steepest intensity scaling with nonlinearity factor $s$ = 14.6 is derived from linear fit of the log-log plot.

After establishing that EMU in ANPs can generate $Eu^{3+}$ emission with high nonlinearity, we sought to corroborate the importance of energy migration within the $Gd^{3+}$ network. To probe energy relay from the sensitizing $Tm^{3+}$ core to the $Eu^{3+}$ activator shell, we prepared a series of core/multishell $NaGdF_4$: 20 mol% $Tm^{3+}/NaY_{1-x}Gd_xF_4/NaGdF_4$: 15 mol% $Eu^{3+}/NaYF_4$ ANPs in which the amount of $Gd^{3+}$ in the intermediate buffer shell was varied ($x$ = 0.0, 0.2, 0.4, 1.0) (**Supporting Information S4**). When no $Gd^{3+}$ was present in the buffer shell ($x$ = 0), energy transfer was suppressed and $Tm^{3+}$ visible upconversion dominated the spectrum (**Figure 3a**, gray spectrum). When we introduced $Gd^{3+}$ ions into the buffer shell ($x$ = 0.2-1.0) $Eu^{3+}$ upconversion (at 510 nm: $^5D_2 \rightarrow {}^7F_3$) could be observed. The intensity of this $Eu^{3+}$ emission line increased relative to that of $Tm^{3+}$ (at 450 nm: $^1D_2 \rightarrow {}^3F_4$) with increasing $Gd^{3+}$ concentration, with the maximum $Eu^{3+}$ emission intensity observed for the case of a $NaGdF_4$ buffer (**Figure 3a**, red spectrum). We were also able to measure steeply nonlinear $Eu^{3+}$ emission in $NaGdF_4$: 20 mol% $Tm^{3+}/NaGdF_4$: 15 mol% $Eu^{3+}/NaYF_4$ ANPs without the intermediate buffer shell (**Supporting Information S5**); however, two orders of magnitude greater laser power was required (grey data points in **Figure 3b**). The above observations reinforce the importance of the $NaGdF_4$ buffer shell for preventing direct $Tm^{3+}$-$Eu^{3+}$ energy cross-talk while preserving EMU. As a further benefit, the buffer shell provides additional passivation of $Tm^{3+}$ core from surface quenching.[20,21]

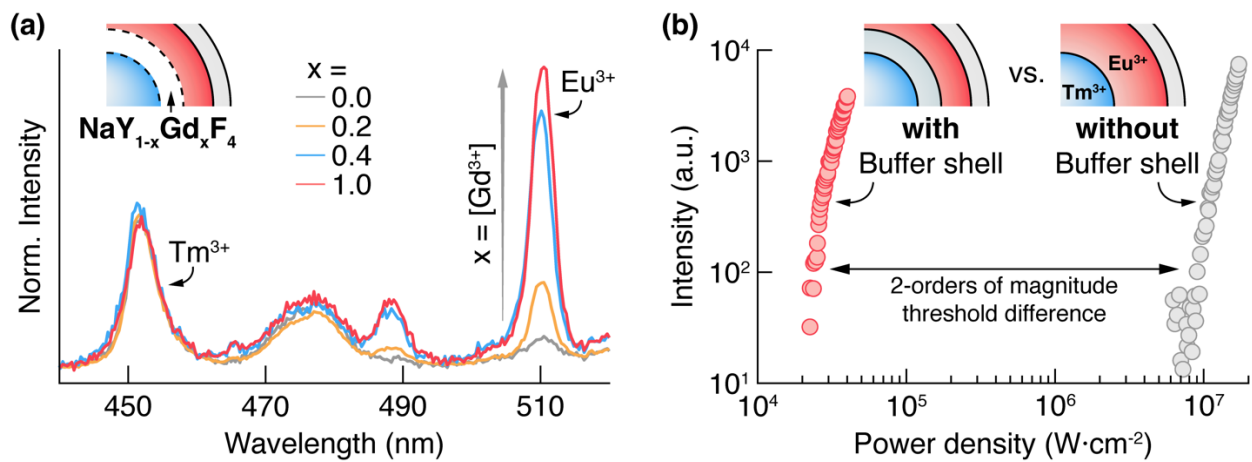

**Figure 3**. (a) Normalized upconversion of $Eu^{3+}$-activated ANPs with varying amounts of $Gd^{3+}$ in the $NaY_{1-x}Gd_xF_4$ buffer shell ($x$ = 0.0, 0.2, 0.4, 1.0). Spectra were acquired with 190 kW·cm$^{-2}$ laser excitation power density to accentuate visible $Tm^{3+}$ upconversion. (b) Pump power dependence of $Eu^{3+}$ upconversion under 1064 nm excitation in ANPs with (red circles) and without (grey

circles) NaGdF$_4$ buffer shell. Two orders of magnitude greater pump powers are required to promote nonlinear Eu$^{3+}$ emission in ANPs without the buffer shell.

Following the demonstration of remote PA in Eu$^{3+}$-activated ANPs, we sought to extend this strategy to other A$^{3+}$ ions to provide a generalized approach for spectral tuning of PA. We prepared a library of ANPs with Tm$^{3+}$-doped cores and activator shells doped with either Tb$^{3+}$, Er$^{3+}$, Ho$^{3+}$, Nd$^{3+}$ or Dy$^{3+}$ ions. These core/multishell ANPs share the same NaGdF$_4$:Tm$^{3+}$ sensitizer core, NaGdF$_4$ intermediate buffer shell, and NaYF$_4$ outer protective shell (as in the Eu$^{3+}$-activated ANPs shown in Figure 1a) – only the composition of the activator (NaGdF$_4$:A$^{3+}$) shell is changed. The above ANPs were observed to have pure β-phase and diameters smaller than 25 nm (**Supporting Information S6**). Under 1064 nm excitation, each composition of ANPs featured a unique spectral profile of the selected A$^{3+}$ activator in the visible spectrum (**Figure 4a, Figure S25**). We note that due to low upconversion intensity and unforeseen contamination by Er$^{3+}$ ions, Nd$^{3+}$ and Dy$^{3+}$-activated ANPs were omitted from further studies. To validate that upconversion in spectrally discrete ANPs stems from PA and EMU, we irradiated Tm$^{3+}$-free NaGdF$_4$: 2 mol% Er$^{3+}$/NaYF$_4$ control nanoparticles with 1064 nm laser (**Supporting Information S7**). Er$^{3+}$ upconversion was only detected at laser power densities greater than 1 MW·cm$^{-2}$, and its power dependence ($s$ = 1.8) followed a two-photon process (**Figure S28b**). In contrast, emissions of different A$^{3+}$ ions in spectrally discrete ANPs were generated in a highly nonlinear fashion, with nonlinearity factors $s$ of 17.2, 11.1, and 10.7 observed for Tb$^{3+}$, Ho$^{3+}$, and Er$^{3+}$-activated ANPs, respectively (**Figure 4b**).

Notably, power densities as low as 6 and 8 kW·cm$^{-2}$ were enough to promote visible PA upconversion from Er$^{3+}$ and Ho$^{3+}$-activated ANPs, respectively, similar to those originally reported for NaYF$_4$:Tm$^{3+}$/NaYF$_4$:Gd$^{3+}$ ANPs[2] and substantially lower than other ANPs.[5] In generating A$^{3+}$ emission across spectrally discrete ANPs we observed some variation of the excitation power density threshold, being the lowest for Er$^{3+}$-activated ANPs and the highest for Tb$^{3+}$-activated ANPs (> 120 kW·cm$^{-2}$, **Figure 4b**). Although the spectrally discrete ANPs were prepared from the same batch of Tm$^{3+}$-doped cores, we found no correlation between Tm$^{3+}$ PA emission (800 nm) power thresholds (**Figure S29**) and that of respective A$^{3+}$ ions; thus, the threshold variability cannot be attributed directly to the sensitizing core. We speculate that these discrepancies arise from small fluctuations in the thicknesses and growth inhomogeneities of buffer NaGdF$_4$ and protective NaYF$_4$ shells. The PA power threshold increases when energy can be directly passed between Tm$^{3+}$ and A$^{3+}$ ions (**Figure 3b**), as well as being highly sensitive to surface quenchers.[21] Additionally, powers above 100 kW·cm$^{-2}$ can lead to photodarkening of Tm$^{3+}$-based ANPs,[8] which further complicates measurements of PA emission from $Tm^{3+}$-activated ANPs. Notwithstanding the power threshold differences, we can generalize that EMU spectral tuning can be adopted in ANPs for generating highly nonlinear emission ($s > 10$) from virtually any activator ion by the $Tm^{3+} \rightarrow Gd^{3+} \rightarrow A^{3+}$ energy transfer cascade (**Figure 1b**, mechanism of $Tm^{3+}$ excitation to $^1I_6$ was informed by numerical simulations – see **Supporting Information S9** for details).

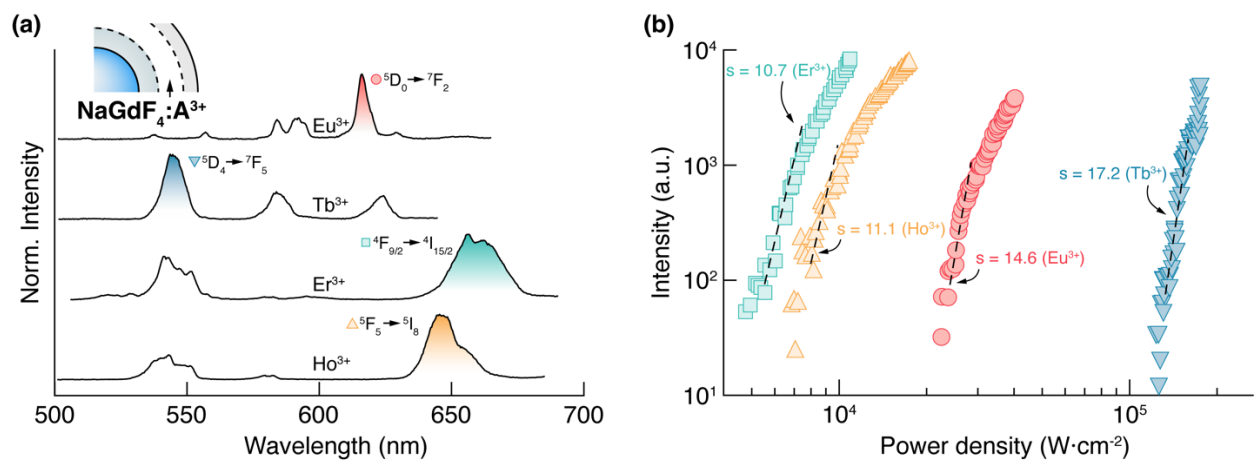

**Figure 4**. (a) Upconversion emission spectra in the 500-700 nm range of spectrally discrete ANPs containing different $A^{3+}$ ions ($Eu^{3+}$, $Tb^{3+}$, $Er^{3+}$ or $Ho^{3+}$) in the activator shell under 1064 nm excitation. Spectra of $Eu^{3+}$, $Er^{3+}$, and $Ho^{3+}$-activated ANPs were taken at 49 kW·cm$^{-2}$ laser power density, that of $Tb^{3+}$-activated ANPs at 580 kW·cm$^{-2}$. (b) Pump power dependence of $A^{3+}$ emission [shaded bands in (a)] intensities under 1064 nm excitation. Intensity scaling with nonlinearity factors s = 10-17 is derived from linear fit of the log-log plot.

Inspired by the above results, we sought to demonstrate how EMU can facilitate the transfer of PA to external fluorophores. To demonstrate this concept, we optically characterized core/shell $NaGdF_4$: 20 mol% $Tm^{3+}/NaGdF_4$ ANPs and core/multishell CdS/CdSe/CdS quantum dots (QDs) co-deposited on a glass coverslip (see **Supplementary Information S10** for details).[22] Under 1064 nm excitation, photoluminescence of both ANPs and QDs could be directly observed (**Figure 5a**). ANP emission stemmed primarily from the $^3H_4 \rightarrow ^3H_6$ transition of $Tm^{3+}$ ions at 800 nm, and a weaker band at 690 nm ($^3F_{2,3} \rightarrow ^3H_6$) could be observed at high excitation powers. Importantly, this $Tm^{3+}$ emission had minimal overlap with the QD emission observed at 630 nm (**Figure S35b**). Although the QDs alone could be excited with 1064 nm laser via two-photon excitation (TPE; **Figure S35c**), in the presence of ANPs their emission intensity increased beyond the quadratic scaling observed without ANPs (**Figure S36a**). The excitation of QDs by ANP → QD energy transfer resulted in QD emission power dependence with s = 10.5 (purple data points in **Figure 5b**; background signal of QD photoluminescence by direct 1064 nm excitation was subtracted from raw data prior to log-log plot fitting). The PA emission of ANPs (blue data points in **Figure 5b**) in the ANP+QD avalanching complex had lower power threshold and higher degree of nonlinearity ($s$ = 15.8), as expected since energy transfer to QDs requires ANP excitation into high energy states and introduces additional relaxation pathways. These experiments corroborate the involvement of $Tm^{3+}$ → $Gd^{3+}$ → QDs energy transfer cascade at around 70 kW·cm$^{-2}$ (for the present combination of nanoparticles).

To further justify QD activation by $Gd^{3+}$ → QDs energy transfer, we tested $NaGdF_4$: 20% $Tm^{3+}$/$NaYF_4$ ANPs as energy donors for QDs, in which case the energy transfer from ANPs to QDs was expected to be minimized by the undoped $NaYF_4$ shell. We observed a slight deviation from the quadratic TPE power-scaling of QD emission in the mixture with $NaGdF_4$:$Tm^{3+}$/$NaYF_4$ ANPs above 100 kW·cm$^{-2}$, however the degree of nonlinearity was significantly lower ($s$ = 6.7, **Figure S36b**). Since both ANPs (with $NaGdF_4$ or $NaYF_4$ outer shell) show near identical $Tm^{3+}$-PA (measured by 800 nm emission), we hypothesize that radiative energy-transfer from $NaGdF_4$:$Tm^{3+}$/$NaYF_4$ ANPs to QDs at high excitation powers can still promote a more nonlinear QD photoluminescence. Altogether, we successfully extended PA from ANPs to other fluorophores via interparticle EMU, demonstrating the ability to form avalanching complexes for highly sensitive multiplexing and sub-diffraction imaging assays.[23,24]

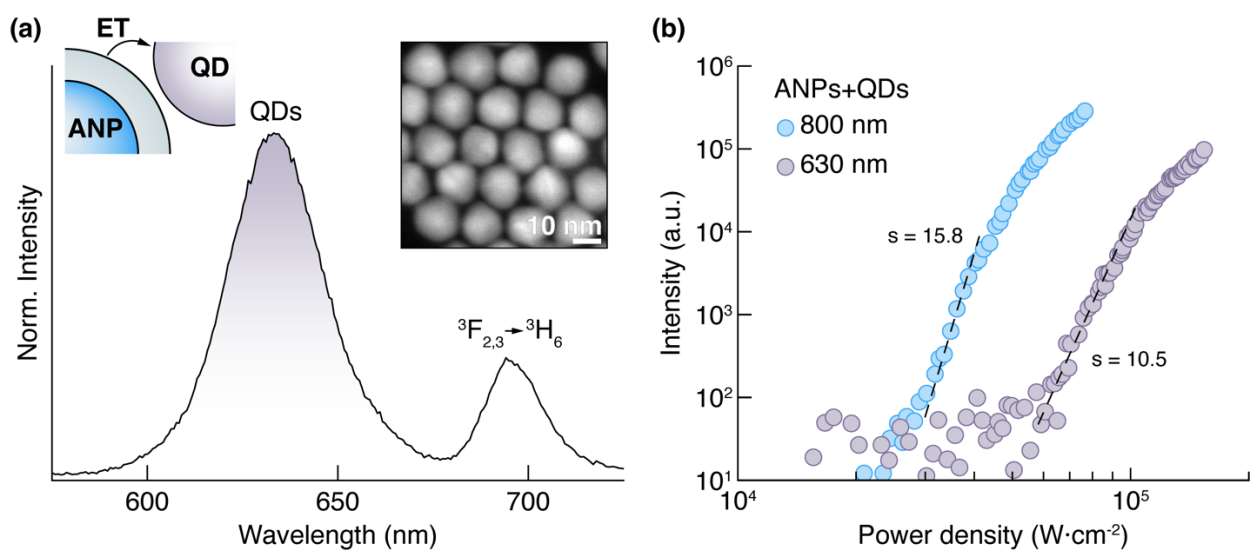

**Figure 5**. (a) Upconversion emission spectra of co-deposited $NaGdF_4$:$Tm^{3+}$/$NaGdF_4$ ANPs and CdS/CdSe/CdS QDs under 1064 nm excitation (100 kW·cm$^{-2}$). ET – energy transfer. Inset: HAADF-STEM image of QDs; scale bar = 10 nm. (b) Pump power dependence of ANP ($Tm^{3+}$ at 800 nm; blue data points) and QD (630 nm as highlighted in (a); purple data points) emission

intensities under 1064 nm excitation. The steepest intensity scaling and corresponding nonlinearity factors (shown in the figure) are derived from a linear fit of the log-log plot.

In summary, we demonstrate that the combination of $Tm^{3+}$-based PA and $Gd^{3+}$-assisted EMU is an effective approach to tune the emission spectra of photon avalanching nanoparticles. Using spectrally discrete ANPs, upconversion profiles of $Er^{3+}$, $Ho^{3+}$, $Eu^{3+}$, $Nd^{3+}$, $Dy^{3+}$, and $Tb^{3+}$ were observed under 1064 nm excitation, with highly nonlinear intensity power-scaling ($s$ = 10-17). Through rational design of the heterostructured ANPs, PA emission of activators was generated in nanocrystals as small as 25 nm in diameter and at exceptionally low laser excitation power densities (<10 kW·cm$^{-2}$). Importantly, we found that PA can be extended beyond lanthanide ions, as demonstrated by PA-like emission of semiconductor QDs in ANP+QD avalanching complexes – which showcases how extreme nonlinearities can be imprinted onto linear emitters. We believe that these findings represent a significant step forward in developing ANPs with customizable compositions, tunable photoluminescence properties, and synergistic interaction with other fluorophores – with potential application in multicolor sub-diffraction imaging, surface patterning, and ultra-sensitive bioassays.

**Methods**
Detailed information regarding the nanoparticle synthesis, structural and spectroscopic characterization is provided in the Supporting Information.

**Associated content**
Supporting Information. Complete ANP and QD synthesis and characterization description. Additional XRD, TEM and photoluminescence data; including Scheme 1, Figures S1-S36, Tables S1-S4 (PDF).

**Notes**
The authors declare no competing financial interest.

**Acknowledgements**
A.S. acknowledges support from the European Union's Horizon 2020 research and innovation program under the Marie Skłodowska-Curie Grant Agreement No. 895809 (MONOCLE). Work at the Molecular Foundry was supported by the U.S. Department of Energy (DOE) under Contract

No. DE-AC02-05CH11231 through the Office of Science, Office of Basic Energy Sciences and J. A. through the Chemical Sciences, Geosciences, and Biosciences Division, Separations Program. H.Y. and quantum dot synthesis were supported by the DOE Office of Energy Efficiency and Renewable Energy (EERE), under Award Number DE-EE0007628. X.Q., B.E.C., and E.M.C. were supported by the Defense Advanced Research Projects Agency (DARPA) under Contract No. HR0011257070. C.L and P.J.S. acknowledge support from DARPA Contract No. HR00112220006 and from the National Science Foundation grant CHE-2203510. This work was financed by the Spanish Ministerio de Innovación y Ciencias under Project No. NANONERV PID2019- 106211RB-I00 and by the Comunidad Autónoma de Madrid (S2022/BMD-7403 RENIM-CM).